\documentclass[aps,pra,notitlepage]{revtex4-1}

\usepackage{graphicx}
\begin{document}

\title{The quantum state should be interpreted statistically}

\author{Holger F. Hofmann}
\email{hofmann@hiroshima-u.ac.jp}
\affiliation{
Graduate School of Advanced Sciences of Matter, Hiroshima University,
Kagamiyama 1-3-1, Higashi Hiroshima 739-8530, Japan}
\affiliation{JST,CREST, Sanbancho 5, Chiyoda-ku, Tokyo 102-0075, Japan
}

\begin{abstract}
In a recent paper (arXiv:1111.3328), Pusey, Barrett and Rudolph claim to prove that statistical interpretations of quantum mechanics do not work. In fact, their proof assumes that all statistical interpretations must be based on hidden variable realism. Effectively, the authors demand from the start that reality must be decided by mathematics, and not by measurements. If this unjustified assumption is dropped, the quantum formalism has a natural statistical interpretation that fully explains the paradox presented by the authors. It is therefore possible to conclude that the paradox actually supports the statistical interpretation, demonstrating once more that quantum mechanics should not be explained by measurement independent realities that are never observed and therefore lie beyond the reach of empirical tests. 
\end{abstract}

\maketitle

The interpretation of quantum mechanics is a tricky business. It is rarely pleasant to live in a state of continuing uncertainty, and it is understandable that theoretical physicists wish to just cut the Gordian knot that keeps us from making clear statements about reality. To me, the recent paper by Pusey, Barrett and Rudolph \cite{PBR} is an example of how such wishful thinking can result in serious errors of judgement. Although the paper also presents a nice little toy model that certainly has merit as a useful addition to the cabinet of quantum paradoxes, the analysis of the model does not really support the sweeping conclusions drawn from it. Specifically, the authors merely analyze a hidden variable model, implicitly assuming that any statistical theory of quantum mechanics must also assume a measurement independent reality.

In their interpretation, the authors completely ignore the fact that the original statistical interpretation of quantum mechanics was explained in terms of the measurement dependence (or contextuality) of quantum mechanics. The explanations given by Bohr, Heisenberg, von Neumann and many others all assume that (a) quantum mechanics should be interpreted statistically, and (b) hidden variables do not work. Could it be, that the authors of \cite{PBR} meant this group of physicists, when thay state that ``Some physicists claim that quantum systems do not have physical properties, or that the existence of quantum systems at all is a convenient fiction''? This statement is certainly a drastic misrepresentation of the empirical position that objects can only be known by their observable effects. The authors contrast this by professing a belief in the existence of ``quantum systems - like atoms and photons.'' However, this seems to contradict their own conclusions, since the reality of a quantum state represented by superpositions of atom or photon numbers should be hard to reconcile with the existence of atoms and photons as real objects.

Obviously, the authors of \cite{PBR} are unaware of the difficulties associated with terms like ``existence'', otherwise they would not think that the conviction that objects must ``exist'' in some absolute form could justify their assumption of a measurement independent reality. The positions that collide here are the empirical tradition that assumes that the existence of an object is known by its observable effects (and nothing else) \cite{Hof09}, and the idealist or dogmatic \cite{comment} position that we know about the existence of an object from an authoritative theory. 

To put the record straight, I would like to point out that the conclusions presented by the authors actually support the statistical interpretation of the quantum state, once the empirically unjustified assumption of dogmatic realism is dropped. In fact, the authors provide several arguments in favor of the statistical interpretation - mainly, that the collapse of the wavefunction is fully explained and loses its mystery. Likewise, entanglement loses much of its mystery. I would like to add that the classical limit is also easier to derive within a statistical framework, as shown by the Wigner function and by the obvious analogy between quantum noise and classical thermal noise in electromagnetic radiation, from radio frequencies to optical frequencies. 

A statistical interpretation that does not refer to hidden variable realities can be based on a decomposition of the quantum state into sub-ensembles that correspond to the different possible measurement outcomes \cite{Hof10}. The structure of quantum theory itself provides a simple and straightforward method to achieve this: the quantum state $\hat{\rho}=\mid \psi \rangle \langle \psi \mid$ can be interpreted as a mixture of any possible set of measurement outcomes $\{\mid f \rangle\}$ \cite{Hof10} by the symmetric products of the state and the measurement operators,
\begin{equation}
\label{eq:joint}
\mid \psi \rangle \langle \psi \mid = \sum_f \frac{1}{2}\left(\mid \psi \rangle\langle \psi \mid f \rangle \langle f \mid + \mid f \rangle\langle f \mid \psi \rangle \langle \psi \mid \right).
\end{equation}
Quantum mechanics does not permit a realist interpretation of these terms, because the elements are non-positive. However, all other properties correspond to those of classical statistics \cite{Hof10,Joh07,Hos10}. Moreover, the negative joint probabilities defined by this operator algebra are actually observed in weak measurement experiments and work very well in the resolution of quantum paradoxes \cite{Res04,Lun09,Yok09,Gog11}. Therefore, dropping the artificial assumption of \cite{PBR} that all possible measurement outcomes must be jointly real even if it is possible to prove that they can never be observed jointly results in an elegant and consistent interpretation of quantum mechanics. It is certainly not necessary to abandon the belief in reality to do so - only, one should wisely restrict such a belief to the empirical reality of actual measurement results, and not believe in an inaccessible ghostworld introduced to satisfy an emotional need for absolute knowledge. 

With this correction, it is possible to analyze the paradox presented by the authors. In the basic version, they consider a pair of overlapping states of a two level system, $\mid 0 \rangle$ and $\mid + \rangle$, where $\mid \pm \rangle = (\mid 0 \rangle \pm \mid 1 \rangle)/\sqrt{2}$. According to the statistical interpretation, one would expect that $0$ and $+$ share a common sub-ensemble $(0,+)$. If two systems are prepared in either $0$ or $+$, any of the four combinations will have a sub-ensemble of $(0,+;0,+)$. The authors then construct a measurement basis from four orthogonal entangled states, such that each measurement outcome has a probability of zero for exactly one of the four possible input states. This result can only be reconciled with the common sub-ensemble, if the necessary positive probability contributions are canceled by appropriate negative probability contributions from other sub-ensembles - and this is exactly what happens in the proper statistical analysis.

The effect is easiest to see in the Bloch vector representation of the state. For $0$, the density operator is $(I+Z)/2$, and for $+$ it is $(I+X)/2$. The measurement operators read
\begin{eqnarray}
\mid \eta_1 \rangle\langle \eta_1 \mid &=& (II+XX-ZZ+YY)
\nonumber \\
\mid \eta_2 \rangle\langle \eta_2 \mid &=& (II+XZ-ZX-YY)
\nonumber \\
\mid \eta_3 \rangle\langle \eta_3 \mid &=& (II-XZ+ZX-YY)
\nonumber \\
\mid \eta_4 \rangle\langle \eta_4 \mid &=& (II-XX+ZZ+YY)
\end{eqnarray}
In this representation, the negative signs of the terms with $X$ and/or $Z$ decide which input combination has zero output probability. For instance, $\eta=1$ has zero probability for an input of $(0,0)$ because of the negative contribution from $ZZ$.
It is now possible to interpret the operators of the initial states as statistical mixtures according to Eq.(\ref{eq:joint}). The components describing the simultaneous assignments of eigenvalues to $X$ and $Z$ read 
\begin{eqnarray}
\hat{R}(0+)&=&I+X+Z
\nonumber \\
\hat{R}(0-)&=&I-X+Z
\nonumber \\
\hat{R}(1+)&=&I+X-Z
\end{eqnarray}
The input $0$ is then given by an equal mixture of $(0+)$ and $(0-)$, and the input $1$ is an equal mixture of $(0+)$ and $(1+)$. In close analogy to classical statistics, the output probabilities for each of the four input combinations have four contributions, one of which is always the common sub-ensemble $\hat{R}(0+)\otimes\hat{R}(0+)$. For the input state $\mid 0,0\rangle$, the contributions to the output probability by sub-ensemble read
\begin{eqnarray}
(\langle \eta=i \mid \hat{R}(0+;0+) \mid \eta_i \rangle)_{i=(1,2,3,4)} &=& (\hspace{0.3cm}1/4, \hspace{0.3cm}1/4,\hspace{0.3cm} 1/4,\hspace{0.3cm} 1/4)
\nonumber \\
(\langle \eta=i \mid \hat{R}(0+;0-) \mid \eta_i \rangle)_{i=(1,2,3,4)} &=& (-1/4, \hspace{0.3cm}3/4, -1/4, \hspace{0.3cm}3/4)
\nonumber \\
(\langle \eta=i \mid \hat{R}(0-;0+) \mid \eta_i \rangle)_{i=(1,2,3,4)} &=& (-1/4, -1/4, \hspace{0.3cm}3/4,\hspace{0.3cm} 3/4)
\nonumber \\
(\langle \eta=i \mid \hat{R}(0-;0-) \mid \eta_i \rangle)_{i=(1,2,3,4)} &=& (\hspace{0.3cm}1/4, \hspace{0.3cm}1/4, \hspace{0.3cm}1/4,\hspace{0.3cm} 1/4)
\end{eqnarray}
The probability of zero for $\eta_1$ therefore originates from the negative joint probabilities for $(0+;0-)$ and for $(0-;0+)$, while the joint probability for the common sub-ensemble $(0+,0+)$ is positive and equally distributed over all outcomes. The statistical interpretation of the density operator $\hat{\rho}$ thus provides a more detailed explanation of the output probabilities - as opposed to a merely formal belief in the reality of the state.

The calculation above shows that the statistical interpretation is ideally suited for the linear algebra of quantum mechanical operators. It is therefore natural and straightforward to interpret the density operator $\hat{\rho}$ of a quantum state as a probability distribution. Of course some physicists may feel uneasy about the appearance of negative probabilities. However, it should be kept in mind that such negative probabilities never describe negative frequencies, since they always refer to joint probabilities of outcomes that cannot be obtained jointly. Thus negative probabilities can explain the origin of the uncertainty principle and provide a more detailed microscopic analysis of quantum physics. On the other hand, it seems that the interpretation of the quantum state as a new reality separate from the directly observed empirical evidence does not lead to any new conclusions. Could it be, that it is actually motivated by a desire to declare the present status of quantum research as complete and final? I see a bit of a danger here that the authors of \cite{PBR} might inadvertently promote an unthinking application of formalism - a dogmatic literalism of equations that would discourage the development of a useful intuitive understanding of the relation between mathematical description and experimental evidence. 

In conclusion, the paradox presented in \cite{PBR} can also be understood as an endorsement of negative joint probabilities for properties that cannot be measured jointly. This interpretation is more consistent with the physics described by the quantum formalism and manages to explain the actual measurement outcomes without the difficulties caused by real collapses and the associated non-localities \cite{Hof11}. Oppositely, it is a bit difficult to see any merit in the dogmatic declaration that the wavefunction is ``real'' and ``exists'', since such an assumption seems to have no consequence for the world of our experience. 

Hopefully, the explanations presented here will clear up the most serious misunderstandings regarding the empirical position that only the outcomes of measurements are definitely real. Specifically, the rejection of measurement independent realities is based on an embrace of experimental reality. It is therefore not an unmotivated opposition to reality as such, but rather a well justified criticism of realities that are introduced for the sole purpose of artificially filling a perceived gap in our understanding. Empirical realism does attribute reality to quantum systems such as atoms and photons and is hence closer to an everyday understanding of reality than the rejection of atoms and photons in favor of an unobservable ``state''. It only requires that such realities must be based on the actual experimental appearance of the objects, and not on a blind and uncritical faith in their mathematical description. The statistical interpretation of quantum mechanics therefore rests on a much sounder scientific foundation than the alternative endorsed by the authors of \cite{PBR}.


\begin{thebibliography}{xyz00}
\bibitem{comment}
By dogmatic, I do not mean to imply that the authors are narrow minded - merely that their beliefs are not based on accessible evidence. Nevertheless, I feel I should point out the similarity to strongly held religious beliefs and the attached doctrines. The danger of dogmatism in science may be greater than physicists would like to think.
\end{thebibliography}
\end{document}